\documentclass[conference]{IEEEtran}
\IEEEoverridecommandlockouts
\usepackage{cite}
\usepackage{amsmath,amssymb,amsfonts}
\usepackage{algorithmic}
\usepackage{graphicx}
\usepackage{subfigure}
\usepackage{multirow}
\usepackage{array}
\usepackage{threeparttable} 
\usepackage{tablefootnote} 

\usepackage{textcomp}
\usepackage{xcolor}
\def\BibTeX{{\rm B\kern-.05em{\sc i\kern-.025em b}\kern-.08em
    T\kern-.1667em\lower.7ex\hbox{E}\kern-.125emX}}
\begin{document}

\title{A Flexible Precision Scaling Deep Neural Network Accelerator with Efficient Weight Combination}

 \author{\IEEEauthorblockN{Liang Zhao\textsuperscript{1,*}, Kunming Shao\textsuperscript{2,3,*}, Fengshi Tian\textsuperscript{2,3}, Tim Kwang-Ting Cheng\textsuperscript{2,3}, Chi-Ying Tsui\textsuperscript{2,3}, Yi Zou\textsuperscript{1,†}}

 \thanks{*Both authors contributed equally.}

 \thanks{†The corresponding author is Yi Zou (e-mail: zouyi@scut.edu.cn)}

 \IEEEauthorblockA{\textsuperscript{1}South China University of Technology, Guangzhou, China\\
 \textsuperscript{2}The Hong Kong University of Science and Technology, Hong Kong SAR, China\\
 \textsuperscript{3}AI Chip Center for Emerging Smart Systems (ACCESS), Hong Kong SAR, China
 }
 
\vspace{-10mm}
}

\maketitle

\begin{abstract}
Deploying mixed-precision neural networks on edge devices is friendly to hardware resources and power consumption. To support fully mixed-precision neural network inference, it is necessary to design flexible hardware accelerators for continuous varying precision operations. However, the previous works have issues on hardware utilization and overhead of reconfigurable logic. In this paper, we propose an efficient accelerator for 2$\sim$8-bit precision scaling with serial activation input and parallel weight preloaded. First, we set two loading modes for the weight operands and decompose the weight into the corresponding bitwidths, which extends the weight precision support efficiently. Then, to improve hardware utilization of low-precision operations, we design the architecture that performs bit-serial MAC operation with systolic dataflow, and the partial sums are combined spatially. Furthermore, we designed an efficient carry save adder tree supporting both signed and unsigned number summation across rows. The experiment result shows that the proposed accelerator, synthesized with TSMC 28nm CMOS technology, achieves peak throughput of 4.09TOPS and peak energy efficiency of 68.94TOPS/W at 2/2-bit operations.

\end{abstract}






\section{Introduction}
The increase in model parameters and size of deep neural network (DNN) poses challenges for hardware computing. On edge devices with constraints on hardware resources and power, quantizing the model has become an effective method to address this issue. To reduce parameter bitwidths without significantly impacting accuracy, some studies \cite{dong2019hawq,wang2019haq} have explored methods for mixed-precision quantization of models. Using fixed-bitwidths accelerators for mixed-precision neural network accelerators will waste hardware resources for low-precision computations. To efficiently support mixed-precision neural network inference, many accelerators are designed with reconfigurable computing units, enabling optimal utilization of hardware resources under different precision computations. 

For the DNN accelerator, the primary computation involves multiply-accumulate (MAC) operations of weights and activations. Depending on the input format of weights/activations and the scaling method of MAC units, mixed-precision accelerators can be classified into bit-parallel and bit-serial architectures, representing precision scaling in spatial and temporal dimensions\cite{camus2019review}. Based on bit-parallel architecture, \cite{moons201714} scales precision by gating larger computing units. However, only hardware resources on the diagonal of the array are used for operations so the hardware utilization is low. Some other works \cite{BitFusion,ryu2022bitblade,li2023precision} achieve precision scaling by combining computing units, which improves hardware utilization, but case more overhead of the reconfigurable logic. 

The accelerators based on bit-parallel architecture mentioned above only support precision with power-of-2 bitwidths, lacking precision flexibility. In contrast, bit-serial accelerators can efficiently achieve continuous precision scaling for serial input operands. \cite{judd2016stripes,albericio2017bit,sharify2018loom} support 1$\sim$16-bit precision with serial input for both weights and activations but they require many cycles for MAC operations and have complex control logic. Some works \cite{lee2018unpu,yang2020bitsystolic,chi202116} are designed with only one of the operands are serial-input. However, their parallel-input operands cannot efficiently support flexible precision scaling. They have problems with hardware utilization when performing low-precision(e.g.,2$\sim$5-bit) operations.

In order to support flexible precision scaling and improve hardware utilization in low-precision computation, we propose an precision-scalable architecture with an efficiency weight combination method. Our contributions are as follows:
\begin{itemize}
\item We set two loading modes for weight operands, 2-bit mode and 3-bit mode. Based on these two modes, we decomposed the weight and the architecture can efficiently support flexible scaling of weight precision.
\item We use systolic dataflow to process activations fed in serial and weights preloaded in parallel. Partial sums computed by the decomposed weights from each column are combined spatially, which improves the hardware utilization in low-precision computations. 
\item We use carry save adder (CSA) tree to sum the multiplication results, with two independent adder tree paths to support both signed and unsigned operations. Compared with the binary adder tree (BAT), the CSA tree in this paper has lower power consumption, especially when processing unsigned numbers.
\end{itemize}
\section{Motivation and Related Work}
For parallel operands, \cite{BitFusion} achieved precision scaling by combining 2bit$\times$2bit computing units. The advantage of this kind of design is that when computing low-precision values, the operation scale increases and the hardware utilization is high. However, the bit-serial design has better precision scaling flexibility and simpler basic unit with shared shift operations than the bit-parallel design.

\begin{figure}[t]
\centering
\includegraphics[width=\columnwidth]{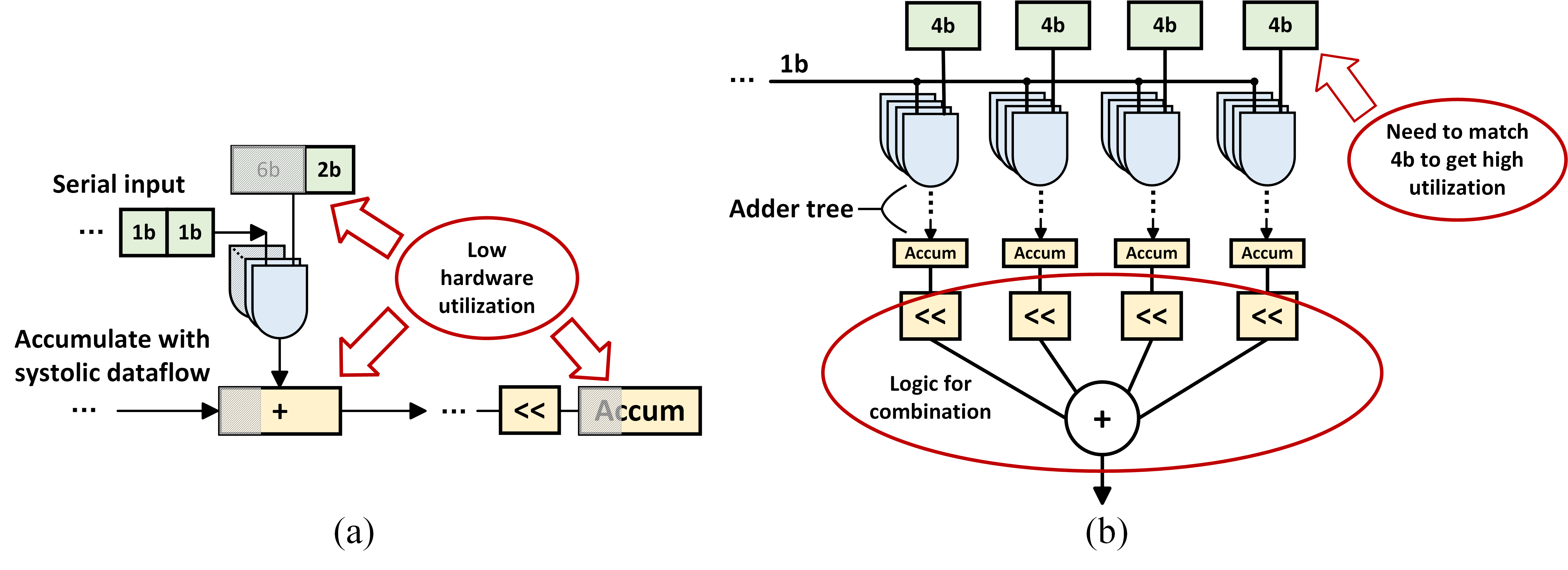}
 \caption{Architecture of (a) \cite{yang2020bitsystolic} and (b) \cite{chi202116}.} 
 \label{relatedwork}
\end{figure}

In the previous bit-serial work, some work \cite{judd2016stripes,sharify2018loom} used serial input for both weights and activations, while others\cite{lee2018unpu,yang2020bitsystolic} used serial input for only one of them. The latter has fewer operation cycles and simpler control logic but the precision scaling of its parallel operands will lead to hardware utilization issues. The accelerator can have high hardware utilization when the parallel operands are input with the bitwidths matching their fixed-size registers because the low-precision operations can only be achieved by gating unnecessary bits in the registers. For example, \cite{yang2020bitsystolic} uses 8-bit register files to store the bit-vector of weight, which means when the weight precision is 2-bit, 6-bit registers are not used and a part of the bitwidths of the adder and accumulator are wasted, as illustrated in Fig. \ref{relatedwork}(a). In addition, \cite{yang2020bitsystolic} does not involve the processing of signed bit. Fig. \ref{relatedwork}(b) shows that \cite{chi202116} achieves  4/8/16-bit weight bitwidths by combining 4-bit units, which is similar to the methods of combining parallel operands in \cite{BitFusion}. We could use lower bitwidths units for combination to increase the precision scaling flexibility and the hardware utilization under the low-precision operations, but the redundancy caused by the combination logic and the appropriate method for scaling precision should be taken into consideration.
\section{Proposed Precision-Scalable Architecture}
\begin{figure}[t]
     \centering
     \includegraphics[width =\columnwidth]{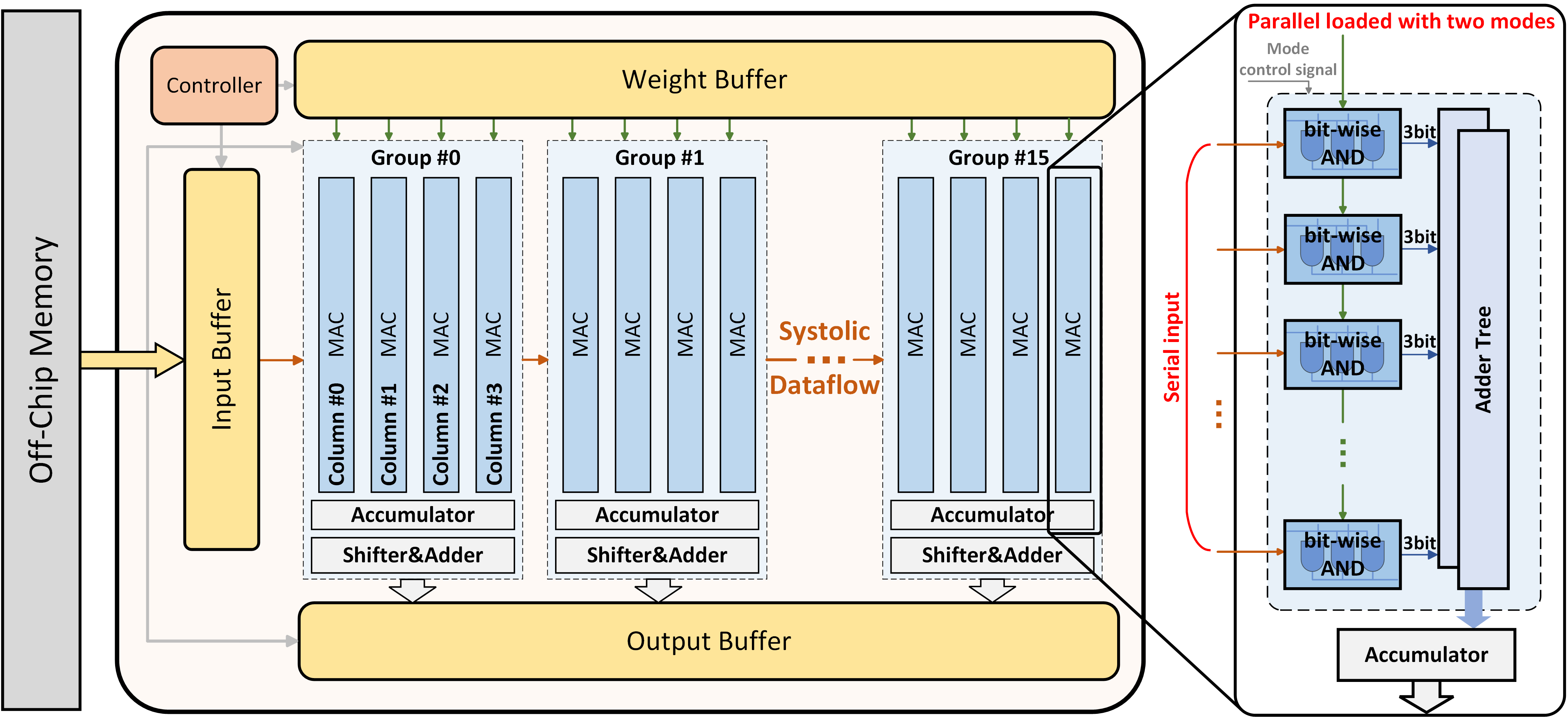}%
     \caption{The overall architecture.}
     \label{fig1}
\end{figure}

To support the bitwidths flexibility of mixed-precision DNNs and improve the hardware utilization under low-precision computations, we propose an efficient precision-scalable architecture, as shown in Fig. \ref{fig1}, with off-chip memory, peripheral buffer and PE array that has 64 rows and 64 columns. We use systolic dataflow so the activations are iterated to each group through registers, which results in a smaller input fanout and allows the accelerator to operate at a higher frequency than multicast dataflow.
\begin{figure}[h]
\centering
\includegraphics[width=0.9\columnwidth]{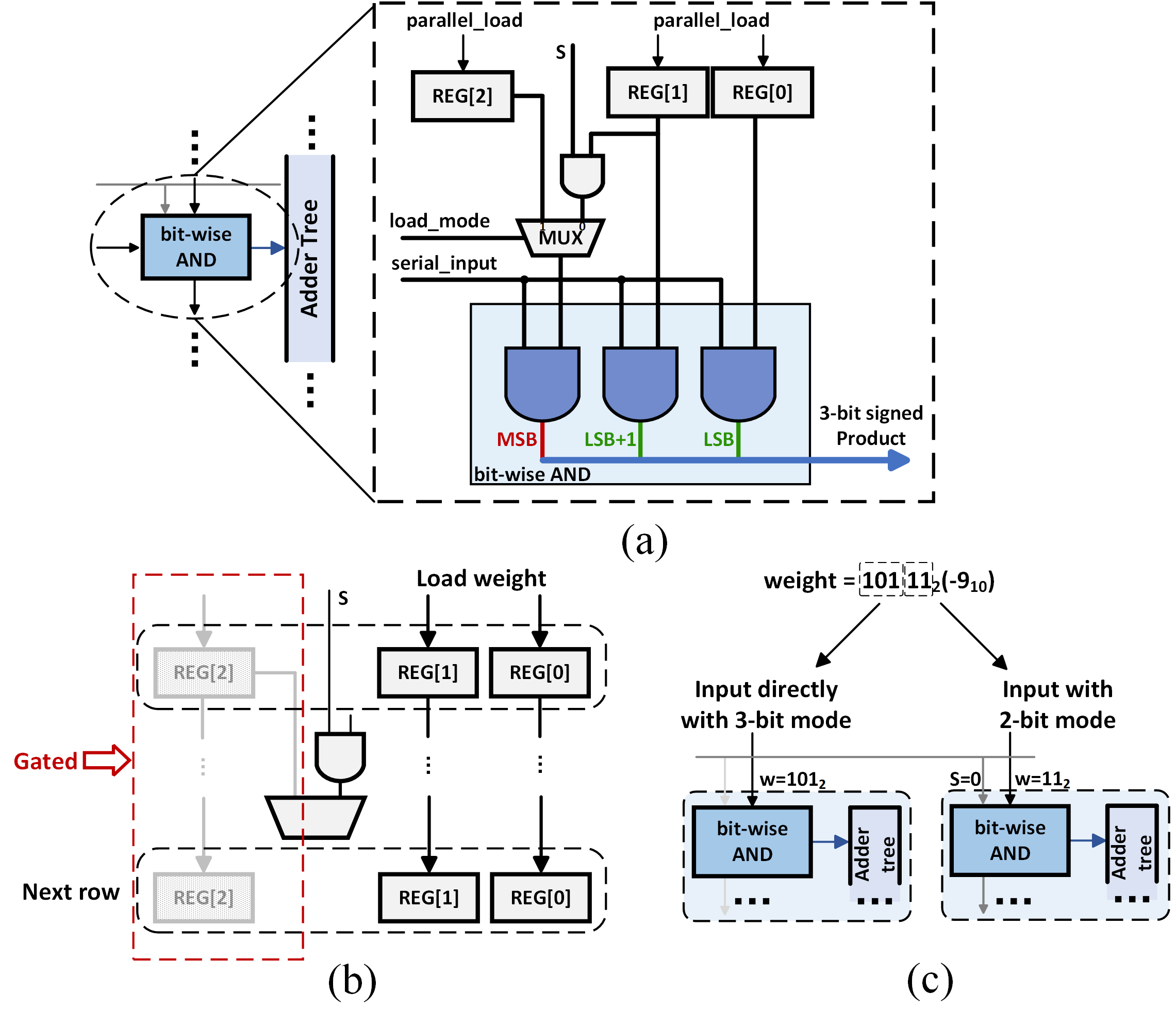}
 \caption{(a) Schematic of the multiplier. (b) Load weight in 2-bit mode. (c) Decompose 5-bit weight with 3-bit and 2-bit mode.} 
 \label{loadw}
\end{figure}
\subsection{Weight Precision Scaling Method}
We design two weight loading modes for each column, 2-bit mode and 3-bit mode. In 2-bit mode, the decomposed weights are input to the multipliers with 2-bit and the extended sign bit is obtained through the signal S which indicates whether the inputs are signed or unsigned. If the input is a signed number, S$=$1, otherwise S$=$0. As shown in Fig. \ref{loadw}(a), the highest bit of the multiplier input is connected to the extended sign bit and the multipliers in each column share the same S signal. Fig. \ref{loadw}(b) illustrates how the weights are loaded in 2-bit mode. Fig. \ref{loadw}(c) shows the way of decomposing and loading weight with 3-bit mode. The highest 3 bits of the weight, including the original sign bit, are directly input to the multipliers and signal S can be ignored. Since the products of the multiplier in both modes are 3-bit signed numbers, the computation logic can be fully reused.

We set four columns as a group for weight combination to minimize the overhead of shifters and the logic for reconfigurability. When all columns in the group are in 2-bit mode, it can support four 2-bit, two 4-bit, one 6-bit and one 8-bit weight operations. Based on this condition, we can set the specific column to 3-bit mode to support four 3-bit, two 5-bit, and one 7-bit weight operations. The bitwidths of decomposed weight are shown in Table. \ref{table1}. 

\begin{figure}[t]
     \centering
     \includegraphics[width =0.8\columnwidth]{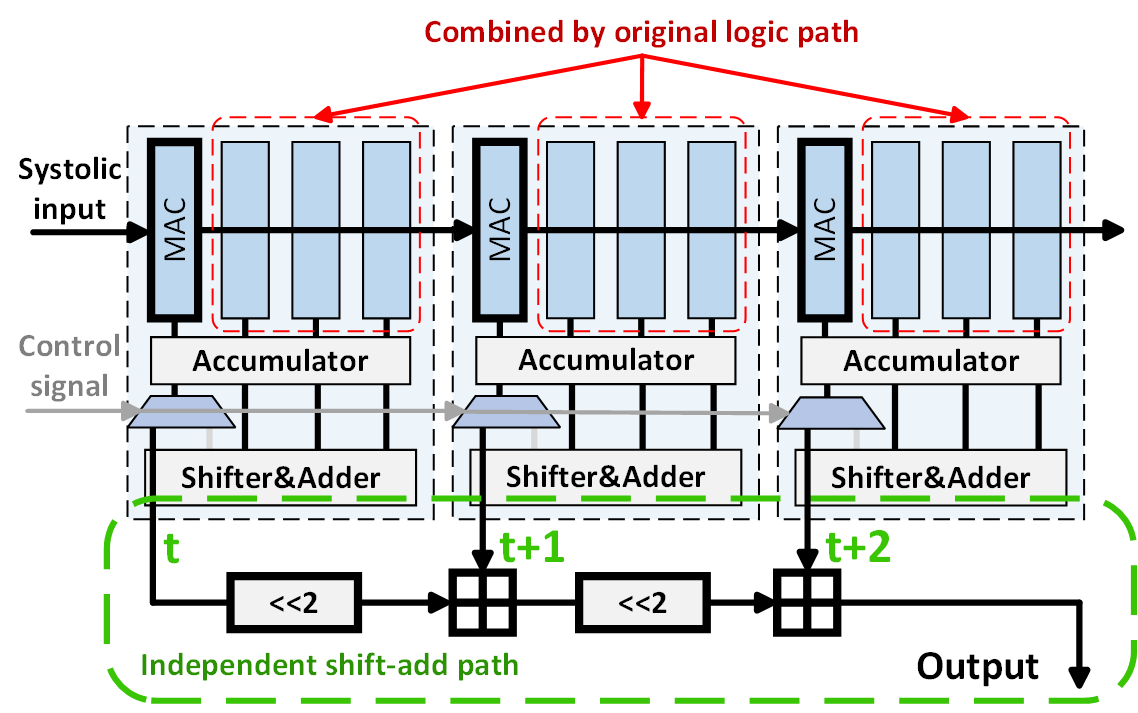}%
     \caption{Independent shift-add path for summation in 6/7-bit weight operation.}
     \label{fig3}
\end{figure}

When the weight precision is 6/7-bit, one column in each group is idle, which can be gated to reduce energy consumption. However, this method reduces array utilization. To fully utilize array resources, we design independent shift-add paths to accumulate the results of these columns. Fig. \ref{fig3} illustrates the shift-add path design based on systolic dataflow and it has the same number of register stages as the original path. Five shift-add paths need to be added, which can be gated in other precision modes, and the whole array only has one column that remains idle. The weights need to be rearranged and routed to the corresponding columns.

Compared to gating large registers or combining weight with the extended sign bit in the previous work, our weight scaling method has higher hardware utilization for low-precision operation and uses the bitwidths occupied by the extended sign bit to achieve higher precision flexibility.
\begin{table}[t]
\centering
\caption{The configure information of decomposed weight and shift-add logic}
\begin{tabular}{|c|ccccccc|}
\hline
\multirow{2}{*}{\textbf{Config Info}} & \multicolumn{7}{c|}{\textbf{Weight Precision}}                                                     \\ \cline{2-8} 
 &
  \multicolumn{1}{c|}{\textbf{8}} &
  \multicolumn{1}{c|}{\textbf{7}} &
  \multicolumn{1}{c|}{\textbf{6}} &
  \multicolumn{1}{c|}{\textbf{5}} &
  \multicolumn{1}{c|}{\textbf{4}} &
  \multicolumn{1}{c|}{\textbf{3}} &
  \textbf{2} \\ \hline
\textbf{Decomposed bit} &
  \multicolumn{1}{c|}{2-2-2-2} &
  \multicolumn{1}{c|}{3-2-2} &
  \multicolumn{1}{c|}{2-2-2} &
  \multicolumn{1}{c|}{3-2} &
  \multicolumn{1}{c|}{2-2} &
  \multicolumn{1}{c|}{3} &
  2 \\ \hline
\textbf{Shifter \#0 (bit)}                   & \multicolumn{1}{c|}{2}  & \multicolumn{2}{c|}{0} & \multicolumn{2}{c|}{2} & \multicolumn{2}{c|}{0} \\ \hline
\textbf{Shifter \#1 (bit)}                   & \multicolumn{5}{c|}{2}                                                    & \multicolumn{2}{c|}{0} \\ \hline
\textbf{Shifter \#2 (bit)}                   & \multicolumn{3}{c|}{4}                           & \multicolumn{4}{c|}{0}                          \\ \hline
\end{tabular}
\label{table1}
\end{table}
\begin{figure}[t]
\centering
\includegraphics[width=0.9\columnwidth]{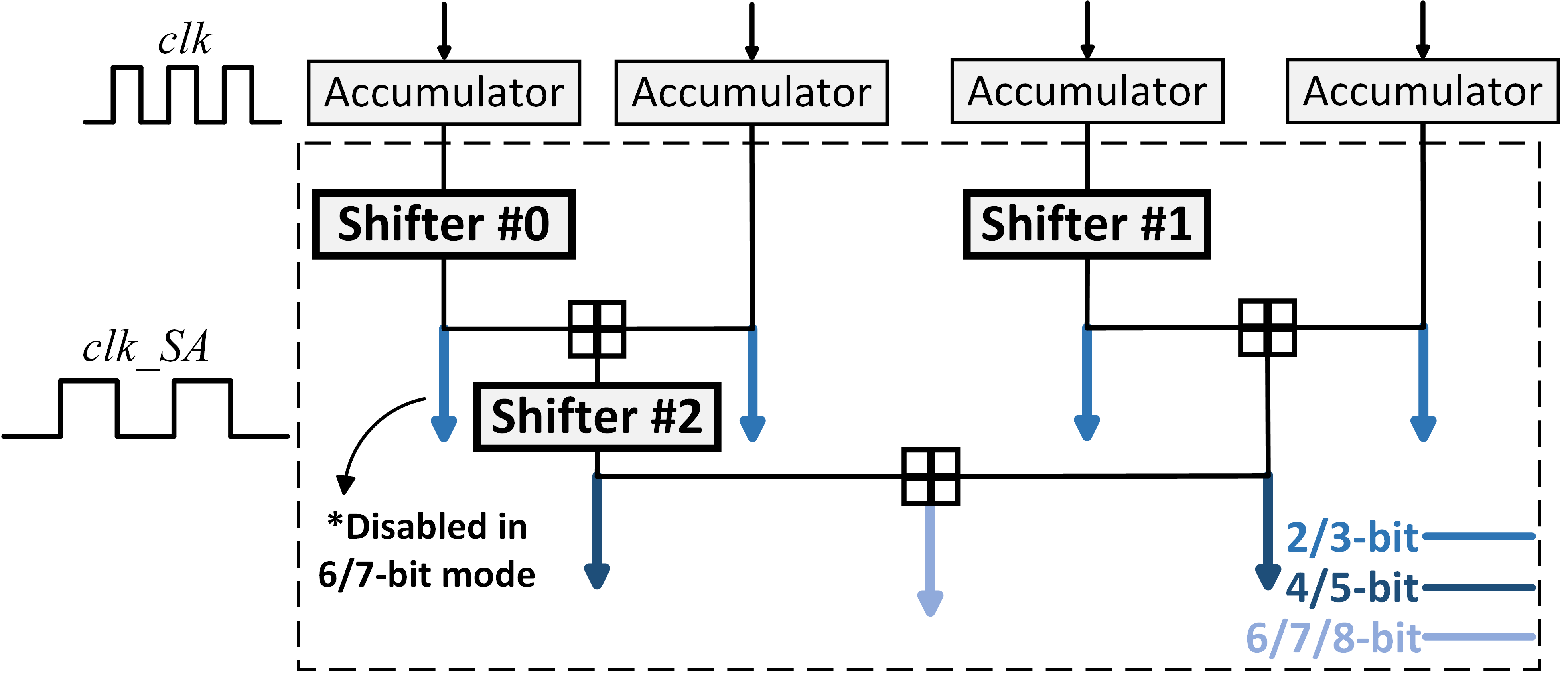}
 \caption{Schematic of configurable shift-add logic.} \label{fig4}
\end{figure}
\subsection{Bit-seial MAC Operation}
The proposed architecture uses weight-stationary systolic dataflow. Weights are preloaded into the array from top to bottom, and activations are fed to each groups by 1-bit iterations. The operands in our work are all represented in 2’s complement format. For M-bit weights and N-bit activations, the principle of MAC operation based on the proposed architecture is as follows:
\begin{equation}
\begin{aligned}
MAC = \sum_{c=0}^{\left [ \frac{M}{2} \right ]- 1} \left (  \sum_{t=0}^{N} \sum_{r=0}^{63}A^{r}\left [ t \right ] \cdot W^{r}_{dcp} \left [ c \right ] \right. \\
\cdot \left ( -1 \right )^{SF}  \cdot 2^{t} \left. \right)\cdot 2^{2c}  
 \end{aligned}
\label{eq1}
\end{equation}
Where $W_{dcp}$ represents the value of decomposed weight assigned to each column, $SF$ is the signal which indicates the sign bit of the activation, $t$ represents different cycles in time dimension, $c$ is column index and $r$ is row index. $W_{dcp}$ is considered positive if it is unsigned. From Eq. (\ref{eq1}), it can be noticed that the weight value of the sign bit of activation is $-2^{N-1}$.

In each column of the array, the decomposed weights are bit-wise multiplied with the 1-bit activation values, and the products from each row are summed by the adder tree. The results of the adder tree from N cycles are input to the accumulator to obtain the MAC results of N-bit activation and decomposed weights. From Eq. (\ref{eq1}), we can know that when the 1-bit activation input is the sign bit, the output value of the adder tree needs to be the inverse value before accumulation.
When the 1-bit activation input is the sign bit and $SF$=1, the output of the adder tree needs to be bit-wise inverted and plus 1. When $SF$=0, the output of the adder tree is directly input to the shift-accumulator. In our design, the activations are input starting from the least significant bit (LSB). Finally, the results of each column are shifted and added to obtain the final result, as shown in Fig. \ref{fig4}. We use another clock domain with lower frequency for the shifter and adder logic to reduce the power consumption and the frequency is set according to activation precision. For example, if activation precision is 8-bit, the frequency of $clk\_SA$ is 1/8 of that of $clk$. Table \ref{table1} illustrates the configuration of the corresponding shifters. For each shifter, it only has two configuration cases, which saves the overhead of reconfigurable logic. 
\begin{figure}[b]
     \centering
     \includegraphics[width =0.75\columnwidth]{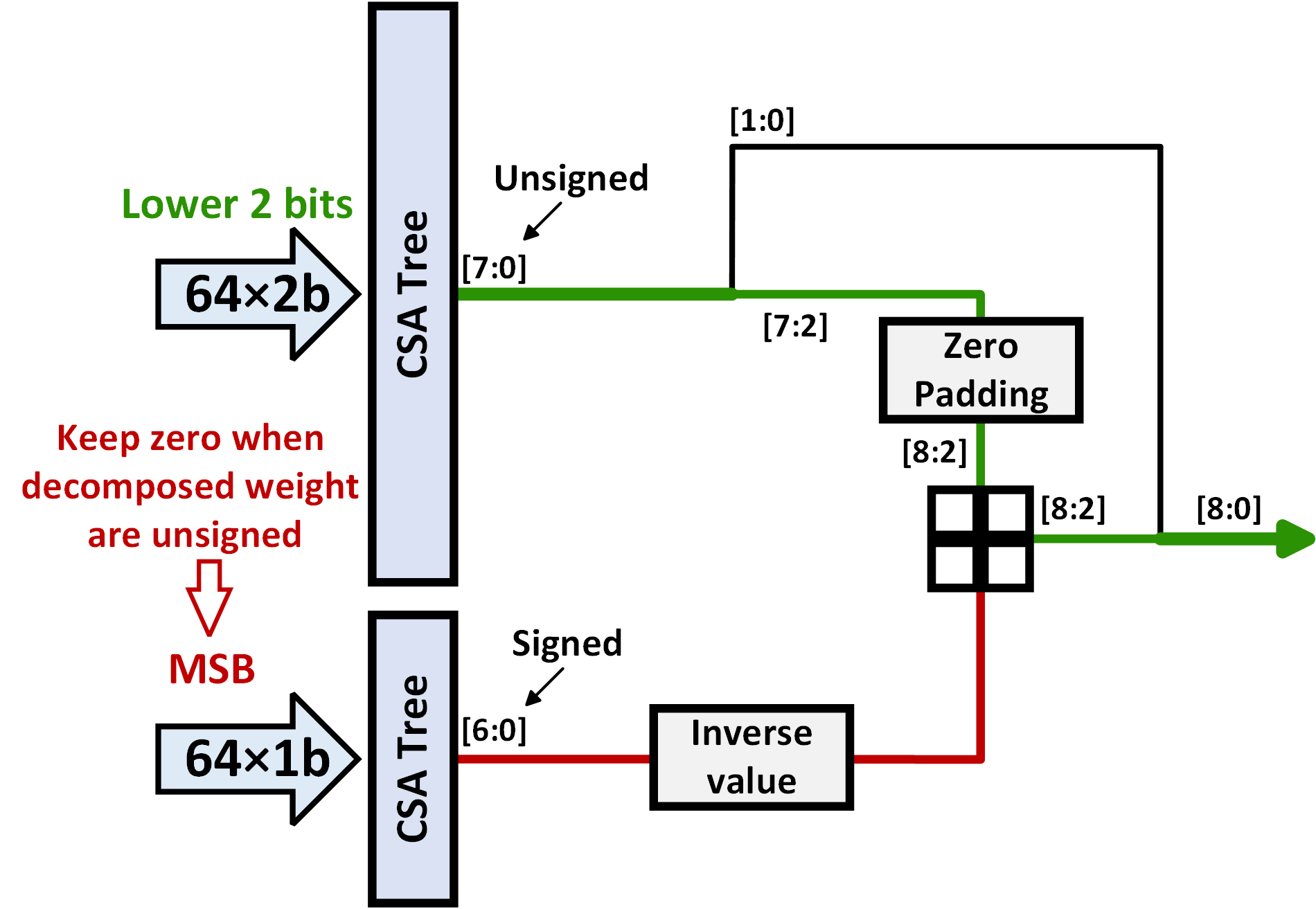}%
     \caption{CSA tree for both signed and unsigned number summation.}
     \label{fig5}
\end{figure}
\subsection{Adder Tree Design}
In this work, we use adder tree consisting of CSA to perform summation operations and optimize the energy efficiency for precision scaling. \cite{sridharan2024ps,shao2024syndcim} has used CSA to design the adder tree which has fewer full adders than BAT. In the proposed work, the adder tree needs to sum 64 3-bit signed numbers. However, as the CSA tree separates the carries and partial sums during the operation, it is hard to take a 3-bit signed number as a whole and process the sign bit as BAT. Thus, we use two independent CSA tree paths to sum the MSB and the lower 2 bits of the products separately, as shown in Fig \ref{fig5}. The result of the MSB adder tree can be considered as the number of 1s among the highest bit of the 64 signed numbers. As mentioned in the last section, the weight value of the sign bit is negative, so the result should be inverse. The lowest 2 bits of the result of lower 2-bit adder tree are directly output, while the highest 7 bits of that are added to the result of the MSB adder tree and then output. When the weights loaded in this column are unsigned numbers, all the inputs of the MSB addition tree are 0. As the two adder tree paths are independent, the number of invalid carries is reduced compared to the BAT.

\section{Experimental Results}
\begin{figure}[h]
     \centering
\includegraphics[width=1\columnwidth]{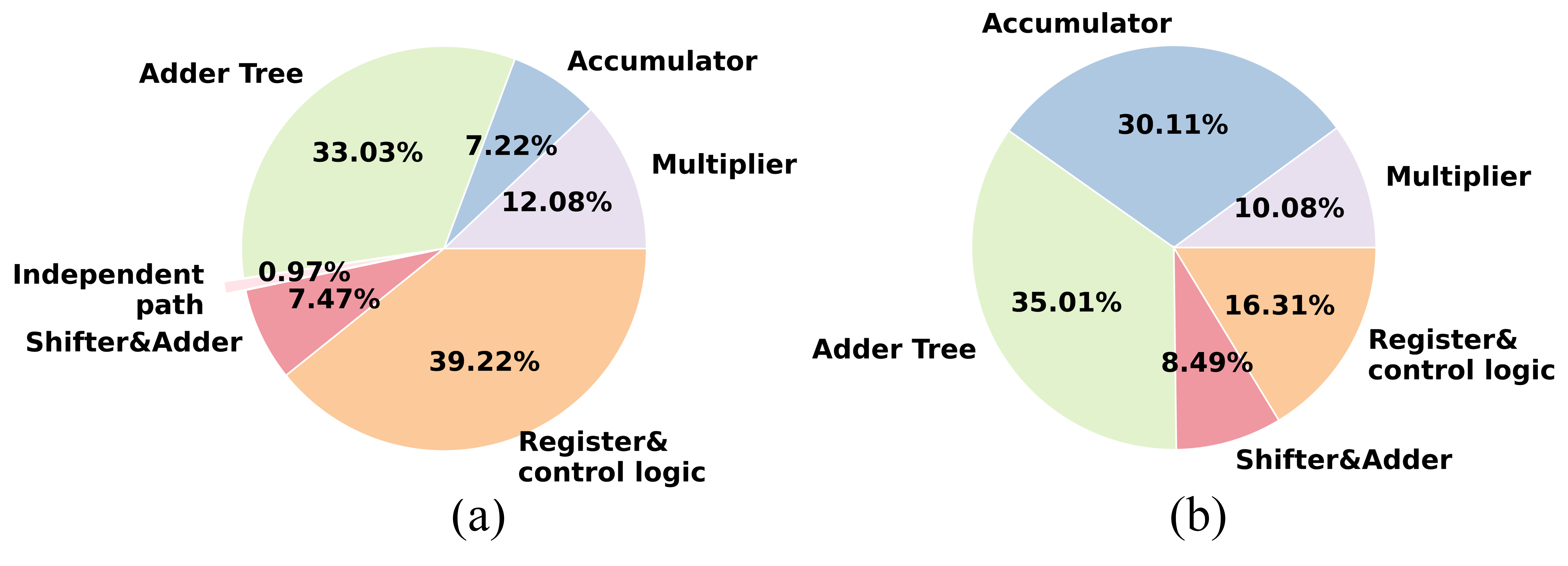}
\vspace{-6mm}
\caption{(a)Area and (b)power breakdown of PE array(8/8-bit).}
\label{fig7}
\end{figure}

We use Design Compiler and PrimeTime PX to evaluate the proposed design and our design is synthesized with TSMC 28nm CMOS technology. We evaluate the area and power breakdown of the PE array, as shown in Fig. \ref{fig7}. The area occupation of the independent shift-add path for 6/7-bit weight operation is only 0.97$\%$, which is considered acceptable. 
 \vspace{-3mm}
 \vspace{-3mm}
 
 We compare the area and power consumption of the adder tree proposed in this paper and the binary adder trees. For the sake of fairness, two types of adder tree are designed and tested with the same input bitwidths and data. Table. \ref{table2} shows the area and power of proposed CSA tree which is normalized according to the BAT. Compared with BAT, our CSA adder tree achieves 15.14$\%$ reduction in area. It has 31.03$\%$ and 22.28$\%$ power reduction when computing unsigned number and signed number respectively.
 \vspace{-4mm}
\begin{table}[h]
\centering
\caption{Area and power consumption of proposed adder tree}
\resizebox{0.8\columnwidth}{!}{
\begin{tabular}{|c|c|cc|}
\hline
\multirow{2}{*}{Normalized value} & \multirow{2}{*}{\textbf{Area}} & \multicolumn{2}{c|}{\textbf{Power}}     \\ \cline{3-4} 
                                  &                                & \multicolumn{1}{c|}{Unsigned} & Signed \\ \hline
\textbf{Proposed CSA Tree}        & 0.8486                         & \multicolumn{1}{c|}{0.6897}    & 0.7772 \\ \hline
\end{tabular}
}
\label{table2}
\end{table}
 \vspace{-2mm}

We measure the energy efficiency of the PE array under different input toggle rates and supply voltages, aiming to demonstrate the general processing efficiency of the PE array. To facilitate testing, we set the bitwidths of weights and activations to be the same, and weight sparsity is set to 50$\%$. The result are shown in Fig. \ref{fig7}, the PE array achieves a peak energy efficiency with a voltage supply of 0.72V under a clock frequency of 500MHz. In this condition, the design has 14TOPS/W, 52.1TOPS/W, 139.8TOPS/W and 205.8TOPS/W energy efficiency at 8/8-bit, 4/4-bit, 3/3-bit and 2/2-bit operations respectively.

\begin{figure}[h]
     \centering
\includegraphics[width=1\columnwidth]{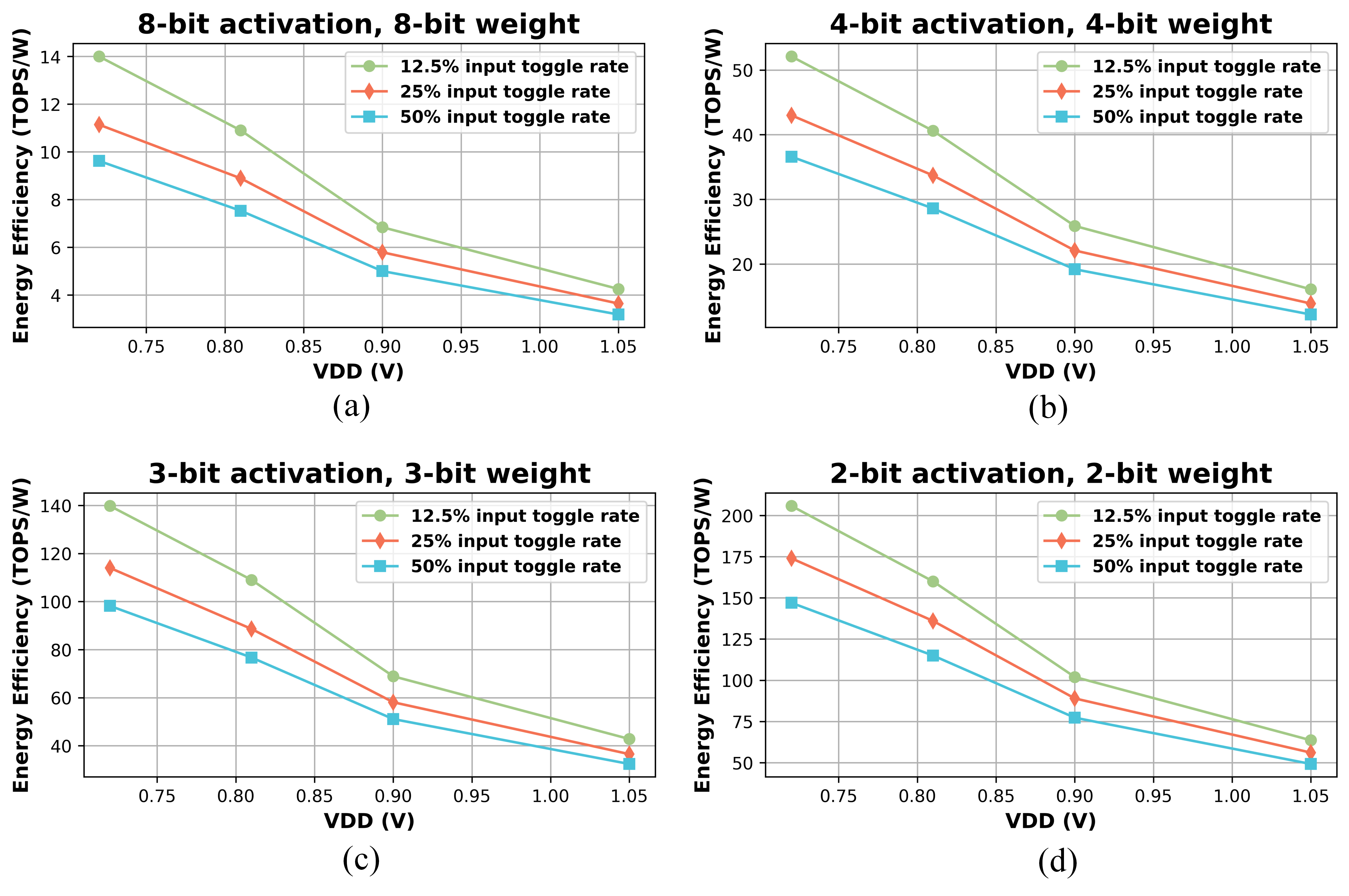}
\vspace{-6mm}
\caption[t]{Energy efficiency of PE array under different input toggle rates.}
\label{fig8}
\end{figure}

The whole accelerator is designed with buffer size of 144KB and tested the inference performance on MobileNetV2\cite{sandler2018mobilenetv2}. Compared to fixed 8-bit model, the accelerator has power reduction of 35.2$\%$ when performing mixed-precision model. We compare the whole flexible precision scaling accelerator with some previous works, as shown in Table \ref{table3}. Compared with the early bit-serial design \cite{lee2018unpu}(UNPU), the proposed work has the advantage of precision scaling flexibility. Our accelerator has higher peak throughput peak energy efficiency in three precision modes than \cite{yang2020bitsystolic}(BitSystolic) which also support flexible precision scaling. The energy efficiency is increased by 18.7$\%$, 10.5$\%$, 11.2$\%$ at 8/8-bit, 4/4-bit, 2/2-bit operations respectively. Compared with the energy-efficiency optimized work based on bit-parallel design\cite{9920733}, the proposed accelerator also performs higher peak energy efficiency, especially at 2/2-bit operations.
\vspace{-4mm}
\begin{table}[h]
\caption{Comparison with other precision-scalable accelerators}
\begin{threeparttable}
\resizebox{\columnwidth}{!}{ 
\begin{tabular}{|c|c|ccc|}
\hline
Type                                                                   & Bit-parallel                                                                    & \multicolumn{3}{c|}{Bit-serial}                                                                                                                                                                                                                                             \\ \hline
Work                                                                   & \begin{tabular}[c]{@{}c@{}}TVLSI'22\\ \cite{9920733}\end{tabular}                    & \multicolumn{1}{c|}{\begin{tabular}[c]{@{}c@{}}JSSC'18\\ \cite{lee2018unpu}$^{\dagger}$\end{tabular}}          & \multicolumn{1}{c|}{\begin{tabular}[c]{@{}c@{}}TCAS-I'20\\ \cite{yang2020bitsystolic}$^{\dagger}$\end{tabular}}                      & \begin{tabular}[c]{@{}c@{}}\textbf{Proposed}\\ \textbf{work}\end{tabular}                          \\ \hline
Tech(nm)                                                                   & 28                                                                            & \multicolumn{1}{c|}{65}                                                                & \multicolumn{1}{c|}{65}                                                                              & 28                                                                             \\ \hline
\begin{tabular}[c]{@{}c@{}}Supply\\ voltage(V)\end{tabular}               & 0.6$\sim$1                                                                             & \multicolumn{1}{c|}{0.63$\sim$1.1}                                                       & \multicolumn{1}{c|}{0.54$\sim$0.66}                                                                    & 0.72$\sim$1.05                                                                   \\ \hline
Freq(MHz)                                                              & 500                                                                          & \multicolumn{1}{c|}{200}                                                              & \multicolumn{1}{c|}{200}                                                                            & 1000$^{1}$                                                                           \\ \hline
Area$^{\ast}$(mm$^{2}$)                                                                   & 1.43                                                                            & \multicolumn{1}{c|}{2.97}                                                               & \multicolumn{1}{c|}{0.74}                                                                            & 0.75                                                                             \\ \hline
Precision                                                              & \begin{tabular}[c]{@{}c@{}}W:2/4/8-bit\\ A:2/4/8-bit\end{tabular}               & \multicolumn{1}{c|}{\begin{tabular}[c]{@{}c@{}}W:1$\sim$16-bit\\ A:16-bit\end{tabular}}  & \multicolumn{1}{c|}{\begin{tabular}[c]{@{}c@{}}W:2$\sim$8-bit\\ A:2$\sim$8-bit\end{tabular}}           & \begin{tabular}[c]{@{}c@{}}W:2$\sim$8-bit\\ A:2$\sim$8-bit\end{tabular}          \\ \hline
\begin{tabular}[c]{@{}c@{}}Peak \\ throughput\\ (TOPS)\end{tabular}   & 4.12                                                                              & \multicolumn{1}{c|}{7.372}                                                               & \multicolumn{1}{c|}{0.403}                                                                             & 4.09                                                                             \\ \hline
\begin{tabular}[c]{@{}c@{}}Peak energy\\ efficiency$^{\ast}$\\ (TOPS/W)\end{tabular} & \begin{tabular}[c]{@{}c@{}}3.62(8-bit)\\ 12.13(4-bit)\\ 22.89(2-bit)\end{tabular} & \multicolumn{1}{c|}{\begin{tabular}[c]{@{}c@{}}7.15(16-bit)\\ 26.93(4-bit)\end{tabular}} & \multicolumn{1}{c|}{\begin{tabular}[c]{@{}c@{}}3.95(8-bit)\\ 15.79(4-bit)\\ 61.98(2-bit)\end{tabular}} & \begin{tabular}[c]{@{}c@{}}4.69(8-bit)$^{2}$\\ 17.45(4-bit)$^{2}$\\ 68.94(2-bit)$^{2}$\end{tabular} \\ \hline
\end{tabular}}
\begin{tablenotes}
\raggedright 
\item[$\ast$]The results are scaled to 28nm.  $^{\dagger}$The works use post-layout estimation. 
\item[1]@1.05V.  $^{2}$@0.72V, 500MHz.
\end{tablenotes}
\end{threeparttable}
\label{table3}
\end{table}
\vspace{-5mm}
\section{Conclusion}
\vspace{-1mm}
This paper proposes a flexible precision scaling accelerator with efficient method of weight combination, supporting 2$\sim$8-bit operations of weights and activations. We first design an efficient weight decomposed method based on two weight loading modes to address the hardware utilization issue under low-precision operation and achieve flexible scaling of weight precision. Then, We designed the architecture with bit-serial MAC operation and configurable shifter and adder logic. Moreover, we design the CSA tree for both signed and unsigned number summation. Experiment results show that proposed accelerator has better peak energy efficiency performance compared to the previous work.

\newpage
\bibliographystyle{IEEEtran} 
\bibliography{IEEEabrv,reference}

\end{document}